\documentclass[prd,showpacs,nofootinbib,twocolumn]{revtex4-1}
\usepackage{amsmath,amsfonts,amssymb}
\usepackage{color}
\usepackage{bm}
\everymath{\displaystyle}

\begin{document}

\title{Manifestations of the rotation and gravity of the Earth in high-energy physics experiments}

\author{Yuri N. Obukhov}
\email{obukhov@ibrae.ac.ru}
\affiliation{Theoretical Physics Laboratory, Nuclear Safety Institute,
Russian Academy of Sciences, B.Tulskaya 52, 115191 Moscow, Russia}

\author{Alexander J. Silenko}
\email{alsilenko@mail.ru}
\affiliation{Research Institute for Nuclear Problems, Belarusian State University,
Bobruiskaya 11, Minsk 220030, Belarus\\
Bogoliubov Laboratory of Theoretical Physics, Joint Institute for Nuclear Research,
141980 Dubna, Russia}

\author{Oleg V. Teryaev}
\email{teryaev@theor.jinr.ru}
\affiliation{Bogoliubov Laboratory of Theoretical Physics, Joint Institute
for Nuclear Research, 141980 Dubna, Russia\\
National Research Nuclear University ``MEPhI'' (Moscow Engineering
Physics Institute), Kashirskoe Shosse 31, 115409 Moscow, Russia}

\date{\today}

\begin {abstract}
The inertial (due to rotation) and the gravitational fields of the Earth affect the motion of an elementary particle and its spin dynamics. This influence is not negligible and should be taken into account in high-energy physics experiments. Earth's influence is manifest in perturbations in the particle motion, in an additional precession of the spin, and in a change of the constitutive tensor of the Maxwell electrodynamics. Bigger corrections are oscillatory and their contributions average to zero. Other corrections due to the inhomogeneity of the inertial field are not oscillatory but they are very small and may be important only for the storage ring electric dipole moment experiments. Earth's gravity causes the Newton-like force, the reaction force provided by a focusing system, and additional torques acting on the spin. However, there are no observable indications of the electromagnetic effects due to Earth's gravity.
\end{abstract}

\maketitle

\section{Introduction}\label{intro}

The high precision of modern experiments makes it necessary to take into account the physical effects due to the Earth's rotation and gravity. The high-energy experiments are performed in a non-Minkowskian spacetime. The off-diagonal components of the metric, $g_{0a} =-(\bm\omega\times\bm r)_{a}/c$, depend on the angular velocity of the Earth's rotation $\omega = 7.27\times10^{-5}$ rad/s, with the average radius vector to a given point on the surface of the Earth $R_\oplus =6378$ km. These metric components are of order of $10^{-6}$ and thus the influence of the Earth's rotation cannot be \emph{a priori} neglected. We also have to take into account the spin precession in the rotating frame and the change of the constitutive tensor of the Maxwell electrodynamics due to the non-Minkowskian spacetime metric. The latter effect is very small, however, the spin motion in the inertial and gravitational fields of the Earth and the influence of the Coriolis, centrifugal, and Newton-like forces on particle's dynamics should be carefully analyzed.

New results obtained in the present work relate to quantum mechanics of a Dirac particle in inertial and electromagnetic fields. We use also the classical approach and pay great attention to a description of spin effects. We perform an analysis of importance of the rotation and gravity of the Earth for contemporary high-energy-physics experiments.

We study the manifestations of the Earth's rotation and gravity in the framework of the relativistic Dirac theory. We demonstrate that, as expected, Earth's rotation gives rise to the Coriolis and centrifugal forces in the laboratory frame. It is also manifested in the additional precession of the spin, as well as in the change of the constitutive tensor of the Maxwell electrodynamics. The latter is encoded in the Riemannian spacetime metric that replaces the flat Minkowski metric. We show that the change of the constitutive tensor of the Maxwell electrodynamics due to the Earth's gravity does not strongly affect particle's dynamics. We analyze the corresponding effects in details. A well-known manifestation of the Earth's rotation is the Sagnac effect \cite{Sagnac}. The electric and magnetic fields acting on the spin in the rotating frame of the Earth coincide with the corresponding fields observed in the inertial (Minkowski) frame comoving an observer carrying the appropriate Schwinger tetrad. The effective electric and magnetic fields defining momentum dynamics and spin motion differ. However, one can see that the larger part of the difference between usual forces and torques in a Minkowski spacetime and the actual forces and torques in Earth's rotating frame vanishes on the average in accelerators and storage rings due to the beam rotation.

Our basic notations and conventions are consistent with the earlier papers \cite{ostrong,ostrong2,ostgrav,ostor,ostor2}. In particular, the world coordinate indices in four dimensions are labeled by the  Latin letters $i,j,k\dots = 0, 1, 2, 3$, whereas we reserve the Greek letters $\mu,\nu, \dots = 0,1,2,3$, for the tetrad indices. The Latin letters from the beginning of the alphabet $a,b,c,\dots = 1,2,3$ denote 3-space world indices. For separate components, we put the hat over anholonomic indices to distinguish them from the holonomic ones. We work with the metric signature $(+, -, -, -)$, and the totally antisymmetric Levi-Civita tensor $\eta_{ijkl}$ has the only nontrivial component $\eta_{0123} = \sqrt{-\det{g_{ij}}}$. We use the system of units with $\hbar=1,~c=1$, although these constants are explicitly displayed in some formulas.

\section{Electromagnetic interactions of a Dirac particle in a rotating frame}\label{rotatframe}

While the classical electrodynamics in curved spacetimes is well understood (see, e.g., Ref. \cite{HehlObukhov}), the quantum mechanics of a Dirac particle in curved spacetime is still a developing subject, see e.g. \cite{Schmutzer1,Schmutzer2,Schmutzer3,Quantmn,Quantmf}. The relativistic quantum-mechanical framework provides the most appropriate description of the dynamics of a spin-${\frac 12}$ particle on an arbitrary spacetime manifold. The perfect agreement of the relativistic quantum mechanics and the classical theory of a spinning particle in arbitrary gravitational and inertial fields has been demonstrated in Ref.~\cite{ostgrav}.

The covariant Dirac equation reads as follows (see Ref. \cite{ostrong2} and references therein)
\begin{equation}
(i\hbar\gamma^\alpha D_\alpha - mc)\Psi=0,\qquad \alpha=0,1,2,3,
\label{Dirac0}
\end{equation}
where $D_\alpha$ is the spinor covariant derivative:
\begin{equation}
D_\alpha = e_\alpha^i D_i,\qquad D_i = \partial _i - {\frac {iq}{\hbar}}
\,A_i + {\frac i4}\sigma^{\alpha\beta}\Gamma_{i\,\alpha\beta}.\label{eqin2}
\end{equation}
This covariant derivative describes the {\it minimal coupling} of a fermion particle with the external classical fields. The list of the latter includes the electromagnetic 4-potential $A_i$ (interacting with the electric charge $q$ of a fermion), and the gravitational field potentials $(e^\alpha_i, \Gamma_i{}^{\alpha\beta})$. The coframe (or tetrad) $e^\alpha_i$ introduces a local Lorentz reference system and is related to the spacetime metric via $g_{ij} = e_i^\alpha e_j^\beta g_{\alpha\beta}$, with the Minkowski metric $g_{\alpha\beta} = {\rm diag}(c^2, -1, -1, -1)$. The Lorentz connection $\Gamma_i{}^{\alpha\beta} = - \Gamma_i{}^{\beta\alpha}$ determines parallel transport; $\sigma^{\alpha\beta} = {\frac i2}\left(\gamma^\alpha \gamma^\beta - \gamma^\beta\gamma^\alpha \right)$ are the generators of the local Lorentz transformations of the spinor field. The flat Dirac matrices $\gamma^\alpha$ are defined in the local Lorentz frames and they have the standard Bjorken-Drell form. Equation (\ref{Dirac0}) describes a Dirac fermion particle minimally coupled to the gravitational (inertial) and the electromagnetic fields.

In Ref. \cite{ostrong2}, we have derived an exact Hermitian Hamiltonian in the Dirac representation corresponding to Eq. (\ref{Dirac0}). In order to give a more detailed description of electromagnetic interactions of a Dirac particle, we should take into account the possible anomalous dipole moments of the particle. Accordingly, we turn to the nonminimal coupling generalization \cite{Itzykson,ostor} of the Dirac equation, introducing the anomalous magnetic moment (AMM) $\mu'$ and the electric dipole moment (EDM) $\delta'$:
\begin{equation}
\left(i\hbar\gamma^\alpha D_\alpha - mc + {\frac{\mu'}{2c}}\sigma^{\alpha\beta}F_{\alpha\beta}
+ {\frac{\delta'}{2}}\sigma^{\alpha\beta}F^*_{\alpha\beta}\right)\Psi=0.
\label{Diracgen}
\end{equation}
The tetrad indices of the Dirac matrices reflect the definition of the three-component physical spin (pseudo)vector in the local Lorentz rest frame of a particle. In the limit of the Minkowski spacetime, Eq. (\ref{Diracgen}) coincides with the Dirac-Pauli equation for a particle with the AMM and EDM (see Ref. \cite{RPJ}). The electromagnetic field tensors are defined as $F_{\alpha\beta}=e_\alpha^i e_\beta^j F_{ij}$ and $F^*_{\alpha\beta} =e_\alpha^i e_\beta^j F^*_{ij}$, where $F_{ij} = \partial_i A_j - \partial_j A_i$ is the electromagnetic field strength tensor and its dual is $F^*_{ij}={\frac 12}\eta_{ijkl}F^{kl}$.

The metric of a general spacetime reads
\begin{equation}\label{ds2}
ds^2 = V^2c^2dt^2 - \delta_{\hat{a}\hat{b}}W^{\hat{a}}{}_cW^{\hat{b}}{}_d(dx^c - K^ccdt)(dx^d - K^dcdt),
\end{equation}
where $a,b,c,\dots = 1,2,3$, and $V, K^a, W^{\hat{a}}{}_b$ are arbitrary functions of time $t$ and space $x^a$.

For the rotating reference system, $V = 1$, $W^{\hat{a}}{}_c = \delta^{\hat{a}}_c$, and $\bm{K} = - \bm{\omega}\times\bm{r}/c$. Choosing the Schwinger gauge for the tetrad (see Refs. \cite{ostrong,ostrong2,Warszawa} and references therein),
\begin{equation}\label{coframe}
e^{\hat{0}}_i = \delta^{\hat{0}}_i,\qquad e^{\hat{a}}_i = \delta^{\hat{a}}_i - cK^{\hat{a}}\delta^{\hat{0}}_i,
\end{equation}
we find explicitly for the {\it anholonomic} components of the Maxwell tensor $F_{\alpha\beta}$,
$\bm{\mathfrak{E}} = \{ F_{\widehat{1}\widehat{0}}, F_{\widehat{2}\widehat{0}}, F_{\widehat{3}\widehat{0}} \}$ and $\bm{\mathfrak{B}} = \{ F_{\widehat{2}\widehat{3}}, F_{\widehat{3}\widehat{1}}, F_{\widehat{1}\widehat{2}} \}$:
\begin{equation}
\bm{\mathfrak{E}} = \bm{E} - (\bm\omega\times\bm{r})\times\bm B,\qquad
\bm{\mathfrak{B}} = \bm{B}, \label{EB}
\end{equation}
The {\it holonomic} components $F_{ij} = \partial_i A_j - \partial_j A_i$ are denoted $\bm{E} = \{ F_{10}, F_{20}, F_{30} \}$ and $\bm{B} = \{ F_{23}, F_{31}, F_{12} \}$. They have the standard form:
\begin{equation}
\bm{E} = -\,\nabla\Phi-\frac{\partial\bm A}{\partial t},\quad \bm{B}=\nabla\times\bm A,\label{eqG}
\end{equation}
with $A_0 = - \Phi$, and ${\bm A} = \{ A_1, A_2, A_3 \}$. The quantities $\bm{\mathfrak{E}}$ and $\bm{\mathfrak{B}}$ are the effective fields in the rotating frame. The anholonomic components of the dual tensor are $F^{*\widehat{a}\widehat{0}} = \mathfrak{B}^a/c$ and $F^{*\widehat{a}\widehat{b}} = \epsilon^{\widehat{a}\widehat{b}\widehat{c}}\mathfrak{E}_c/c$. The tensors $F_{\alpha\beta}, F^*_{\alpha\beta}$ and the effective fields $\bm{\mathfrak{E}},~\bm{\mathfrak{B}}$ are defined with respect to the local Lorentz Schwinger frame (\ref{coframe}).

The Dirac Hamiltonian found in Ref. \cite{ostgrav} does not contain additional terms characterizing the AMM and EDM. In the case under consideration, the inclusion of these dipole terms leads to the generalized Dirac Hamiltonian:
\begin{eqnarray}
{\cal H} &=& \beta mc^2 + q\Phi + c\bm\alpha\cdot\bm\pi - \bm\omega\cdot(\bm r\times\bm\pi)
-\frac{\hbar}{2}\bm\omega\cdot\bm\Sigma\nonumber\\ \label{HDP}
&& - \,\bm{\Pi}\cdot\bm{\mathcal{M}} - i\bm{\gamma}\cdot\bm{\mathcal{P}},
\end{eqnarray}
where we defined two objects with the dimension of the energy by
\begin{equation}\label{MaPa}
\bm{\mathcal M} = \mu'\bm{\mathfrak{B}} + \delta'\bm{\mathfrak{E}},\qquad
\bm{\mathcal P} = c\delta'\bm{\mathfrak{B}} - \mu'\bm{\mathfrak{E}}/c,
\end{equation}
and $\bm\pi=-i\hbar\bm{\nabla} - q\bm A$ is the kinetic momentum operator.

Now we perform the unitary transformation from the Dirac picture to the Foldy-Wouthuysen (FW) representation \cite{FW} using the method developed in Refs. \cite{PRA,FWproof}. The FW picture holds a special place in the relativistic quantum mechanics due to some unique properties. In this representation, the quantum mechanical operators for relativistic particles in an external field have the same form as in the nonrelativistic quantum theory. In particular, the position operator \cite{NW} and momentum operator are $\bm r$ and $\bm p=-i\hbar\bm{\nabla}$, respectively, and the polarization operator for spin-$1/2$ particles is described by the Dirac matrix $\bm\Pi = \beta\bm{\Sigma}$. In other representations, these operators are expressed by much more complicate formulas (see Refs. \cite{JMP,FW}). The relations between the operators in the FW representation are analogous to the relations between the corresponding classical quantities, and the simple form of the operators corresponding to the classical observables is a great advantage of this representation. The above-mentioned properties of the FW representation make it useful for the description of a transition to the semiclassical approximation and to the classical limit of the relativistic quantum mechanics \cite{FW,CMcK}. It is worthwhile to note that the Hamiltonian and all other operators are diagonal in 2-spinors (block-diagonal) in this representation.

In the FW representation, the transition to the classical limit is usually accomplished by replacing the operators in expressions for the Hamiltonian and in operator equations of motion with the corresponding classical quantities. The possibility of such a replacement, explicitly or implicitly used in practically all works devoted to the relativistic FW transformation, was recently rigorously proven in Ref. \cite{JINRLett12}. This essentially simplifies the interpretation of the basic quantum mechanical equations, especially in the relativistic case.

We can recast the Hamiltonian (\ref{HDP}) into the form
\begin{equation}
{\cal H}=\beta mc^2 + {\cal E}+{\cal O},\qquad\beta{\cal E}={\cal E}\beta,
\qquad\beta{\cal O}=-{\cal O}\beta, \label{2eq3}
\end{equation}
where ${\cal E}$ and ${\cal O}$ are even and odd operators (diagonal and off-diagonal in 2-spinors), respectively.

The resulting transformed Hamiltonian in the FW representation is exact for terms proportional to the zeroth and first powers of the Planck constant and also for terms proportional to $\hbar^2$ and describing contact interactions. It is given by \cite{PRA,FWproof,TMPFW}
\begin{equation}
\begin{array}{c}
{\cal H}_{FW}=\beta\epsilon + {\cal E}-\frac 18\left\{\frac{1}
{\epsilon(\epsilon + mc^2)},\left[{\cal O},\left[{\cal O},{\cal
F}\right]\right]\right\}, \\ {\cal F}={\cal E}-i\frac{\partial}{\partial
t},\qquad\epsilon=\sqrt{m^2c^4  + {\cal O}^2}. \end{array} \label{eq14gen}
\end{equation}
The square brackets ${[}\ ,\ {]}$ and the curly brackets $\{\ ,\ \}$ denote the commutators and anti-commutators, respectively. In the case under consideration, it is convenient to present the FW Hamiltonian in the form \cite{ostor}
\begin{widetext}
\begin{eqnarray}
{\cal H}_{FW} &=& {\cal H}^{(D)}_{FW}+{\cal H}^{(add)}_{FW},\label{gnfFW}\\
{\cal H}^{(D)}_{FW} &=& \beta \epsilon' + q\Phi -\bm\omega\cdot(\bm r\times\bm\pi)
- {\frac{\hbar}{2}}\bm\omega\cdot\bm\Sigma-\frac{q\hbar c^2}{4}\left\{\frac{1}{\epsilon'},
\bm\Pi\cdot\bm B\right\}\nonumber\\
&& +\,\frac{q\hbar c^2}{8}\left\{\frac{1}{\epsilon'(\epsilon' + mc^2)},\Bigl[\bm\Sigma\cdot
(\bm\pi\times\bm{\mathfrak{E}} - \bm{\mathfrak{E}}\times\bm\pi)-\hbar\bm{\nabla}
\cdot\bm{\mathfrak{E}}\Bigr]\right\},\label{FWHe}\\
{\cal H}^{(add)}_{FW} &=& -\,{\frac c4}\biggl\{\frac{1}{\epsilon'},\Bigl[\bm\Sigma\cdot
(\bm\pi\times\bm{\mathcal{P}} - \bm{\mathcal{P}}\times\bm\pi)-\hbar\bm{\nabla}
\cdot\bm{\mathcal{P}}\Bigr]\biggr\} - \bm\Pi\cdot\bm{\mathcal{M}}\nonumber\\
&& +\,{\frac{c^2}4}\biggl\{\frac{1}{\epsilon'(\epsilon' + mc^2)},\Bigl[(\bm\Pi\cdot\bm\pi)(\bm\pi
\cdot\bm{\mathcal{M}})+(\bm{\mathcal{M}}\cdot\bm\pi)(\bm\Pi\cdot\bm\pi) \nonumber\\
&& +\,\beta\frac\hbar2\left(\bm\pi\!\cdot\!\bm{\mathcal{J}} + \bm{\mathcal{J}}
\!\cdot\!\bm\pi\right) - \beta{\frac {\hbar}{2c}}\Bigl\{\bigl([\bm\omega\times\bm r]
\cdot\bm{\nabla}\bigr),(\bm\pi\cdot\bm{\mathcal{P}})\Bigr\}\Bigr]\biggr\},\label{eq3t}
\end{eqnarray}
where $\epsilon'=\sqrt{m^2c^4 + \bm\pi^2c^2}$ and $\bm{\mathcal{J}}
= \bm{\nabla}\times\bm{\mathcal{M}} + {\frac {\partial\bm{\mathcal{P}}}{c\partial t}}$.
\end{widetext}

The operator ${\cal H}^{(D)}_{FW}$ arises from the transformation of the Hamiltonian that corresponds to the Eq. (\ref{Dirac0}), and the operator ${\cal H}^{(add)}_{FW}$ contains the terms, proportional to the AMM and EDM. In the last term of Eq. (\ref{eq3t}), the nabla operator acts on $\bm{\mathcal{P}}$ and defines the derivatives of this quantity.

It is important that Eqs. (\ref{gnfFW})-(\ref{eq3t}) are exact in the above-mentioned sense that the terms proportional to the zeroth, first, and second powers of $\hbar$ are exact. Notice that $\mu', \delta', \bm{\mathcal{P}}$, and $\bm{\mathcal{M}}$ are proportional to $\hbar$.

In the special case of a \emph{nonrelativistic} Dirac particle without the AMM and EDM, the Hamiltonian (\ref{eq3t}) has been obtained in Refs. \cite{Schmutzer1,Schmutzer2,Schmutzer3}. It is compatible with the Hamiltonian for a relativistic Dirac particle with the AMM and EDM which has been derived in Refs. \cite{RPJ,PRA} in the framework of quantum electrodynamics. The Hamiltonian for a relativistic spin-1 particle with the AMM and EDM in electromagnetic fields derived in Ref. \cite{PRDspin1} has a similar form.

If we neglect an influence of the spin on the particle motion, the world velocity operator is given by \cite{PRD2007}
\begin{eqnarray}
\bm{v}\equiv\frac{d\bm r}{dt}=\frac{i}{\hbar}[{\cal H}_{FW},\bm r]= c^2\beta\frac{\bm\pi}{\epsilon'}-\bm\omega\times\bm{r}.\label{velocityo}
\end{eqnarray}
The classical limit is derived by a replacement of the operators with the
corresponding classical quantities \cite{JINRLett12}.

\section{Dynamics of a relativistic particle in external fields}\label{Effield}

The discussion of the (classical and quantum) dynamics of a particle with spin under the action of the external electromagnetic, inertial and gravitational fields has a long  (see, for example, \cite{kuro,dimo,bahlnp,bahyo,Schiff,Heer,Webster,Menegozzi,Gron,Scorgie}) and quite contradictory (different authors often arrived at incompatible conclusions) history. In order to clarify the issue and to analyse the possible effects of the external fields in the high energy physics, we present here a consistent framework for the investigation of particle's dynamics in the physical conditions on the Earth.

Let us summarize what we know about these physical conditions: the Earth is rotating with the angular velocity (taking the duration of a sidereal day $T_\oplus =$ 23 hours 56 minutes 4.1 seconds = 86164.1 s)
\begin{eqnarray}
\omega_\oplus = {\frac {2\pi}{T_\oplus}} = 7.29\times 10^{-5} {\rm s}^{-1},\label{oE}
\end{eqnarray}
and the Earth is heavy, with the mass $M_\oplus = 5.97\times 10^{24}\,$kg. Despite such a mass, on the surface of Earth, that has an average radius $R_\oplus = 6.378\times 10^6\,$m, the gravitational field is quite weak: the corresponding value of gravitational potential is
\begin{equation}
{\frac {GM_\oplus}{c^2R_\oplus}} = 6.95\times 10^{-10}.\label{GMC2R}
\end{equation}
Here $G$ is Newton's gravitational constant and $c$ is the speed of light.
Note the difference between the sidereal day and the solar one ($T=$ 24 hours).

We now need to choose the appropriate spacetime geometry, which correctly describes the terrestrial inertia and gravity, in order to study the relativistic particle dynamics on the Earth. Taking into account the weakness of the gravitational created by a rotating source, we choose the parameters of the spacetime metric (\ref{ds2}) as  $W^{\hat{a}}{}_c = W\delta^{\hat{a}}_c$, and
\begin{eqnarray}
V &=& 1 - {\frac {GM}{c^2r}},\label{VLT}\\
W &=& 1 + {\frac {GM}{c^2r}},\label{WLT}\\
\bm{K} &=& -\,{\frac {\overline{\bm{\omega}}\times\bm{r}}c}.\label{KLT}
\end{eqnarray}
Here $M = M_\oplus$, and
\begin{equation}
\overline{\bm{\omega}} = \bm{\omega} - {\frac {2G}{c^2r^3}}\bm{J}.\label{oJ}
\end{equation}
On the right-hand side, the first term accounts for the rotation of the Earth: $\bm{\omega} = \left(0, 0, \omega_\oplus\right)$, when we (conventionally) choose the rotation axis along the third coordinate. The second term in (\ref{oJ}) is determined by the total angular momentum $\bm{J} = \left(0, 0, J_\oplus\right)$ of the source. In Einstein's general relativity, the metric (\ref{ds2}) with (\ref{VLT})-(\ref{oJ}) describes the Lense-Thirring geometry of a weak gravitational field created by a massive rotating compact object.

Let us compare the two terms in (\ref{oJ}) for the terrestrial conditions. We estimate the angular momentum of the Earth as $J_\oplus = I_\oplus\omega_\oplus = {\frac {2M_\oplus R^2_\oplus}{5}} \omega_\oplus = 7.08\times 10^{33} {\frac {{\rm kg\,m}^2}{\rm s}}$, which is consistent with the value that is usually found in the literature \cite{Ciu}. As a result, in a terrestrial laboratory, we have
\begin{equation}\label{oJLT}
{\frac {2G}{c^2R_\oplus^3}}\,J_\oplus = 4.05\times 10^{-14}\,{\rm s}^{-1}.
\end{equation}
As we see, the second term in (\ref{oJ}) is nine orders smaller than the first one, cf. Eqs. (\ref{oE}) and (\ref{oJLT}). This explains why do we need the constant $\bm{\omega}$ in Eq. (\ref{oJ}): the Lense-Thirring (second term) contribution cannot describe the correct value of the inertial (Coriolis and centrifugal) forces in Earth's labs, as it was noticed already in \cite{dimo}, for example.

With an account of such a huge domination of the first term over the second one, for all practical purposes it is sufficient to put $\overline{\bm{\omega}} = \bm{\omega}$, which from now on we assume in our subsequent computations. Accordingly, the Lense-Thirring geometry is actually reduced to the Schwarzschild metric in rotating coordinates. More precisely, Eqs. (\ref{VLT})-(\ref{KLT}) describe the weak field approximation of the isotropic form of the Schwarzschild metric, with an account for the rotation of the terrestrial laboratory system.

After these preliminaries, we are in a position to study the particle dynamics. The motion of a relativistic test particle with mass $m$ and charge $e$ in the gravitational and electromagnetic fields is described by the generally covariant equation
\begin{equation}\label{Dui}
{\frac {Du^i}{d\tau}} = {\frac {du^i}{d\tau}} + \Gamma_{kj}{}^i\,u^ku^j
= -\,{\frac em}\,g^{ij}F_{jk}u^k.
\end{equation}
The trajectory of a test particle $x^i(\tau)$ is parametrized by the proper time $\tau$, and the 4-velocity vector $u^i = {\frac {dx^i}{d\tau}}$ is normalized by the condition $g_{ij}u^iu^j = c^2$. The 3-velocity $\bm{v} = \{v^a\}$ is defined as usual by $v^a = {\frac {dx^a}{dt}} = {\frac {u^a}{u^0}}$.

For the metric of the Earth, (\ref{ds2}) and (\ref{VLT})-(\ref{KLT}), we find the connection
\begin{eqnarray}\label{GS1}
\Gamma_{00}{}^0 &=& 0,\quad \Gamma_{0a}{}^0 = -\,{\frac {g_a}{c^2}},
\quad \Gamma_{ab}{}^0 = 0,\\ \label{GS2}
\Gamma_{00}{}^a &=& -\,g^a + [\bm{\omega}\times(\bm{\omega}\times\bm{r})]^a,\\
\Gamma_{0b}{}^a &=& \delta^{ae}\epsilon_{ecb}\,\omega^c,\label{GS3}\\
\Gamma_{bc}{}^a &=& {\frac {1}{c^2}}\left(\delta^a_bg_c + \delta^a_cg_b
- \delta_{bc}g^a\right).\label{GS4}
\end{eqnarray}
These formulas are derived, as it was assumed from the very beginning, in the weak field approximation, where we consistently neglect the corrections proportional to (\ref{GMC2R}). Here Newton's acceleration vector is as usual
\begin{equation}
\bm{g} = -\,{\frac {GM}{r^3}}\,\bm{r}.\label{N}
\end{equation}

Substituting (\ref{GS1})-(\ref{GS4}), and recalling the definition of the electric and
magnetic fields (\ref{eqG}), the equations of motion (\ref{Dui}) read, in components:
\begin{eqnarray}
{\frac {du^0}{d\tau}} &=& (u^0)^2\,2{\frac {\bm{g}\cdot\bm{v}}{c^2}}\nonumber\\
&& + \,u^0{\frac {e}{mc^2}}\left(\bm{E} + \bm{v}\times\bm{B}\right)\cdot
\left(\bm{v} + \bm{\omega}\times\bm{r}\right),\label{EduF0}\\
{\frac {d(u^0\bm{v})}{d\tau}} &=& (u^0)^2\left\{\left(1 + {\frac {v^2}{c^2}}
\right)\bm{g} - 2\bm{v}\,{\frac {\bm{g}\cdot\bm{v}}{c^2}}\right\}\nonumber\\
&& -\,(u^0)^2\left\{\bm{\omega}\times(\bm{\omega}
\times\bm{r}) + 2\bm{\omega}\times\bm{v}\right\}\nonumber\\
&& +\,u^0{\frac {e}{m}}\Big[\bm{E} + \bm{v}\times\bm{B} \nonumber\\
&& -\,{\frac {\bm{\omega}\times\bm{r}}{c^2}}\,\left(\bm{E} + \bm{v}\times\bm{B}
\right)\cdot\left(\bm{v} + \bm{\omega}\times\bm{r}\right)\Big].\label{EduFa}
\end{eqnarray}
Using (\ref{EduF0}) in (\ref{EduFa}), we finally derive, after going from the proper
time to the coordinate one with ${\frac 1{u^0}}\,{\frac {d}{d\tau}} = {\frac {d}{dt}}$,
\begin{eqnarray}
{\frac {d\bm{v}}{dt}} &=& \left(1 + {\frac {v^2}{c^2}}
\right)\bm{g} - 4\bm{v}\,{\frac {\bm{g}\cdot\bm{v}}{c^2}}\nonumber\\
&& -\,\bm{\omega}\times(\bm{\omega}\times\bm{r}) - 2\bm{\omega}\times\bm{v}\nonumber\\
&& +\,{\frac {e}{mu^0}}\Big[\bm{E} + \bm{v}\times\bm{B} \nonumber\\
&& -\,{\frac {\bm{v} + \bm{\omega}\times\bm{r}}{c^2}}\,\left(\bm{E} + \bm{v}\times\bm{B}
\right)\cdot\left(\bm{v} + \bm{\omega}\times\bm{r}\right)\Big].\label{Edvdt}
\end{eqnarray}
In the non-relativistic limit (for slow moving particle, with ${\frac {v^2}{c^2}} \ll 1$), we obviously obtain the correct structure of the total force as a sum of Newton's gravitational force $\bm{g}$, the inertial centrifugal and Coriolis forces $-\bm{\omega}\times(\bm{\omega} \times\bm{r}) - 2\bm{\omega}\times\bm{v}$, and the Lorentz force ${\frac {e}{m}} (\bm{E} + \bm{v}\times\bm{B})$.

The above discussion concerns spinless particles. In order to include the spin effects, we need to consider the dynamics with respect to the anholonomic local Lorentz frame. For the metric (\ref{ds2}) and (\ref{VLT})-(\ref{KLT}), we choose the Schwinger tetrad
\begin{equation}\label{coframe1}
e^{\hat{0}}_i = V\delta^{\hat{0}}_i,\qquad e^{\hat{a}}_i = W(\delta^{\hat{a}}_i - cK^{\hat{a}}\delta^{\hat{0}}_i).
\end{equation}
The anholonomic components of the 4-velocity $u^\alpha = e^\alpha_iu^i$ then read
\begin{equation}
u^{\widehat{0}} = Vu^0,\qquad u^{\widehat{a}} = Wu^0(v^a - cK^a),\label{uuLT}
\end{equation}
and hence for the 3-velocity we find
\begin{equation}
\widehat{\bm{v}} = {\frac {u^{\widehat{a}}}{u^{\widehat{0}}}}
= {\frac WV}(\bm{v} + \bm{\omega}\times\bm{r}).\label{vhatLT}
\end{equation}

The {\it anholonomic} components of the Maxwell tensor $F_{\alpha\beta} = e^i_\alpha e^j_\beta F_{ij}$
are related to the holonomic components:
\begin{equation}
\bm{\mathfrak{E}} = {\frac 1{VW}}\{\bm{E} - (\bm\omega\times\bm{r})\times\bm{B}\},\qquad
\bm{\mathfrak{B}} = {\frac 1{W^2}}\bm{B}. \label{EB1}
\end{equation}
For the Earth, in view of (\ref{GMC2R}) the gravitational corrections are extremely small, hence with very good accuracy we have $V = W = 1$. Then (\ref{coframe1}) and (\ref{EB1}) reduce to (\ref{coframe}) and (\ref{EB}), respectively, thus recovering the rotating frame case with the inertial effects only. The relation (\ref{vhatLT}) between the anholonomic and holonomic 3-velocity is then simplified to $\widehat{\bm{v}} = \bm{v} + \bm{\omega}\times\bm{r}$.

The equations of motion in the Lorentz frame look formally similar to (\ref{Dui}):
\begin{equation}\label{eomA}
{\frac {du^\alpha}{d\tau}} + \Gamma_{\gamma\beta}{}^\alpha\,u^\gamma u^\beta
 = -\,{\frac em}\,g^{\alpha\beta}F_{\beta\gamma}u^\gamma.
\end{equation}
However now we have to use the Minkowski metric $g^{\alpha\beta} = e^\alpha_ie^\beta_jg^{ij} = {\rm diag}({\frac 1{c^2}}, -1, -1, -1)$ here, and the components of the local Lorentz connection read
\begin{eqnarray}
\Gamma_{\widehat{0}\widehat{a}}{}^{\widehat{0}} &=& -\,{\frac {g_a}{c^2}},\quad
\Gamma_{\widehat{0}\widehat{0}}{}^{\widehat{a}} = -\,g^a,\label{connLT1}\\ \label{connLT2}
\Gamma_{\widehat{0}\widehat{b}}{}^{\widehat{a}} &=& -\,\delta^{ad}\epsilon_{dbc}\omega^c,\\
\Gamma_{\widehat{c}\widehat{b}}{}^{\widehat{a}} &=& {\frac 1{c^2}}(g_b\delta^a_c
- g^a\delta_{bc}). \label{connLT3}
\end{eqnarray}
These expressions are also approximate, because for the Earth we have (\ref{GMC2R}),
so that we put $V = 1$ and $W = 1$ in the final formulas.

As a result, we obtain the equations of motion (\ref{eomA}) in components
\begin{eqnarray}\label{dgLT}
{\frac {d\gamma}{d\tau}} &=& {\frac {\gamma^2}{c^2}}\,\bm{g}\cdot\widehat{\bm{v}}
+ {\frac e{mc^2}}\,\gamma\,\bm{\mathfrak{E}}\cdot\widehat{\bm{v}},\\
{\frac {d(\gamma\widehat{\bm{v}})}{d\tau}} &=& \gamma^2\Big\{\big(1
+ {\frac {\widehat{v}^2}{c^2}}\big)\bm{g} - \widehat{\bm{v}}\,{\frac {\bm{g}\cdot
\widehat{\bm{v}}}{c^2}} - \bm{\omega}\times\widehat{\bm{v}}\Big\}\nonumber\\
&& +\,{\frac e{m}}\gamma\left(\bm{\mathfrak{E}}
+ \widehat{\bm{v}}\times\bm{\mathfrak{B}}\right).\label{dgvLT}
\end{eqnarray}
Here we denoted $\gamma = u^{\widehat{0}}$. From the normalization of the 4-velocity we can identify this with the Lorentz factor: $\gamma = 1/\sqrt{1 - \widehat{v}^2/c^2}$. Using (\ref{dgLT}) in (\ref{dgvLT}), and switching from the proper time to the coordinate time:
\begin{eqnarray}
{\frac {d\widehat{\bm{v}}}{dt}} &=& \big(1 + {\frac
{\widehat{v}^2}{c^2}}\big)\bm{g} - 2\widehat{\bm{v}}\,{\frac {\bm{g}\cdot
\widehat{\bm{v}}}{c^2}} - \bm{\omega}\times\widehat{\bm{v}}\nonumber\\
&& +\,{\frac e{m\gamma}}\left(\bm{\mathfrak{E}}
+ \widehat{\bm{v}}\times\bm{\mathfrak{B}} - {\frac {\widehat{\bm{v}}}{c^2}}
\bm{\mathfrak{E}}\cdot\widehat{\bm{v}}\right).\label{dvtLTA}
\end{eqnarray}

It is important to notice that the equations of motion (\ref{EduF0})-(\ref{Edvdt}) and (\ref{dgLT})-(\ref{dvtLTA}) are completely equivalent. They describe the dynamics of the relativistic charged particle in the electromagnetic and gravitational fields with respect to the different reference frames. But provided we use the relations between the velocities $\widehat{\bm{v}} = \bm{v} + \bm{\omega}\times\bm{r}$ and the electric and magnetic fields (\ref{EB}), it is straightforward to explicitly recast (\ref{EduF0})-(\ref{Edvdt}) into (\ref{dgLT})-(\ref{dvtLTA}) or the other way round. It is worthwhile to mention that $\bm{E} + \bm{v}\times\bm{B} = \bm{\mathfrak{E}} + \widehat{\bm{v}}\times\bm{\mathfrak{B}}$.

There exists yet another description of the test particles: the Hamiltonian approach. The general Hamiltonian function for an arbitrary spacetime geometry was obtained in \cite{ostgrav}. For the metric of the Earth, (\ref{ds2}) and (\ref{VLT})-(\ref{KLT}), the Hamiltonian reads
\begin{equation}\label{HamLT}
{\mathcal H}(\bm{p}, \bm{r}) = V\sqrt{m^2c^4 + c^2\pi^2/W^2} + c\bm{K}\cdot\bm{\pi} + e\Phi.
\end{equation}
Here the generalized momentum is $\bm{\pi} = \bm{p} - e\bm{A}$, with
$\pi^2 = \delta_{ab}\pi^a\pi^b$.

It is straightforward to derive the canonical equations
\begin{equation}
{\frac {d\bm{r}}{dt}} = {\frac {\partial{\mathcal H}}{\partial\bm{p}}},\qquad
{\frac {d\bm{p}}{dt}} = -\,{\frac {\partial{\mathcal H}}{\partial\bm{r}}}.\label{can}
\end{equation}
The first equation yields a relation between the momentum and velocity:
\begin{equation}
\bm{\pi} = Wm\widehat{\bm{v}}\gamma.\label{piv}
\end{equation}
Remarkably, it turns out that the momentum is directed along the anholonomic velocity (\ref{vhatLT}), and not along the coordinate velocity. The second Hamilton equation (\ref{can}) defines the force $\bm{F} = d\bm{p}/dt$ and describes the dynamics of the momentum:
\begin{equation}
{\frac {d\bm{\pi}}{dt}} = - \,\bm{\omega}\times\bm{\pi}
+ m\bm{g}\gamma\Big(1 + {\frac {\widehat{v}^2}{c^2}}\Big)
+ e(\bm{E} + \bm{v}\times\bm{B}),\label{canLT2}
\end{equation}
Using (\ref{piv}) in (\ref{canLT2}), we recover the equation of motion (\ref{dgvLT}). This proves the equivalence of the Hamiltonian approach with both the holonomic and anholonomic descriptions of particle's dynamics.

The importance of the Hamiltonian picture is based on the fact that the analysis of the quantum dynamics of the Dirac fermion, which we discussed in Sec.~\ref{rotatframe}, leads to the Foldy-Wouthuysen quantum Hamiltonian (its spin-free part) which in the semiclassical limit coincides with the classical Hamiltonian (\ref{HamLT}). Therefore, the study of the quantum dynamics of the spin ${\frac 12}$ in external classical fields for the Earth yields exactly the same results by replacing the quantum operators with the corresponding classical variables.

\section{Spin dynamics in external fields}\label{rttnmnn0}

Motion of a charged test particle with spin and dipole moments in the gravitational and electromagnetic fields is described by the equations for the 4-velocity $u^\alpha$ and the vector of spin $S^\alpha$. The corresponding system consists of the equation (\ref{eomA}) for the velocity, and the equation for the spin:
\begin{eqnarray}
{\frac {dS^\alpha}{d\tau}} + \Gamma_{\gamma\beta}{}^\alpha\,u^\gamma S^\beta =
 -\,{\frac em}\,g^{\alpha\beta}F_{\beta\gamma}S^\gamma \nonumber\\ \label{eomS}
-\,{\frac 2\hbar}\left[M^\alpha{}_\beta + {\frac {1}{c^a}}\left(
M_{\beta\gamma}u^\alpha u^\gamma - M^{\alpha\gamma}u_\beta u_\gamma\right)\right]S^\beta.
\end{eqnarray}
The spin is affected by an additional force proportional to the generalized polarization tensor
\begin{equation}
M_{\alpha\beta} = \mu'\,F_{\alpha\beta} + c\delta'\,F^*_{\alpha\beta}.\label{Mab}
\end{equation}
The anomalous magnetic moment can be written as
\begin{equation}
\mu' = {\frac {e(g - 2)\hbar}{4m}} = {\frac {g - 2}2}\,\mu_B. \label{mu}
\end{equation}
where $g$ is the gyromagnetic factor and $\mu_B$ is Bohr's magneton. The dual electromagnetic tensor $F^*_{\alpha\beta} = {\frac 12}\eta_{\alpha\beta\mu\nu}F^{\mu\nu}$ couples to the electric dipole moment $\delta'$ with the dimension of the charge times length, $[\delta'] = [e\ell]$. Similarly to the magneton, one can introduce a convenient unit of an electric dipole moment. A reasonable definition is the electron charge times the electron Compton length: $\delta_B = e\,{\frac {\hbar}{mc}}$. Then (\ref{mu}) can be extended to both types of the dipole moments:
\begin{equation}
\mu' = a\,{\frac {e\hbar}{2m}},\qquad \delta' = b\,{\frac {e\hbar}{2mc}}, \label{mude}
\end{equation}
The dimensionless constant parameters $a = (g -2)/2$ and $b$ characterize the
magnitude of the anomalous magnetic and electric dipole moments, respectively.

In components, the generalized polarization tensor (\ref{Mab}) reads
\begin{eqnarray}
M_{\widehat{0}\widehat{a}} = c{\mathcal P}_a,\qquad M_{\widehat{a}\widehat{b}} = \epsilon_{abc}
{\mathcal M}^c,\label{MM}
\end{eqnarray}
where $\bm{\mathcal P}$ and $\bm{\mathcal M}$ are defined in (\ref{MaPa}).

Note that the complete set of equations of motion (\ref{eomA}) and (\ref{eomS}) are
written with respect to the anholonomic frame.

Let us now specialize to the physical conditions of the Earth. The gravitational and inertial effects are taken into account in the Lorentz connection coefficients (\ref{connLT1})-(\ref{connLT3}). The resulting system (\ref{eomA}) and (\ref{eomS}) then reads, in components:
\begin{eqnarray}\label{dg0}
{\frac {d\gamma}{d\tau}} &=& {\frac e{mc^2}}\,\gamma\,\bm{\mathfrak{E}}{}_{\rm eff}
\cdot\widehat{\bm{v}},\\ \label{dgv0}
{\frac {d(\gamma\widehat{\bm{v}})}{d\tau}} &=& {\frac e{m}}\gamma\left(
\bm{\mathfrak{E}}{}_{\rm eff} + \widehat{\bm{v}}\times\bm{\mathfrak{B}}{}_{\rm eff}\right),\\
{\frac {dS^{\widehat{0}}}{d\tau}} &=& \bm{S}\cdot\left\{{\frac e{mc^2}}
\,\bm{\mathfrak{E}}{}_{\rm eff} - {\frac {2\gamma^2}{c^2\hbar}}\,\widehat{\bm{v}}
\times\bm{\Delta}\right\},\label{ds0}\\
{\frac {d\bm{S}}{d\tau}} &=& S^{\widehat{0}}\left\{{\frac e{m}}\bm{\mathfrak{E}}{}_{\rm eff}
- {\frac {2\gamma^2}{\hbar}}\,\widehat{\bm{v}}\times\bm{\Delta}\right\}\nonumber\\
&& + \bm{S}\times\left\{{\frac e{m}}\bm{\mathfrak{B}}{}_{\rm eff} + {\frac {2\gamma^2}
{\hbar}}\left(\bm{\Delta} - \widehat{\bm{v}}\,{\frac {\bm{\Delta}\cdot\widehat{\bm{v}}}
{c^2}}\right)\right\}.\label{dgsa}
\end{eqnarray}
Here, as before, $u^\alpha = \{\gamma, \gamma\widehat{\bm{v}}\}$, and we have for the spin components $S^\alpha = \{S^{\widehat{0}}, \bm{S}\}$. The first two equations (\ref{dg0}) and (\ref{dgv0}) are equivalent to the equations of motion (\ref{dgLT}) and (\ref{dgvLT}), where we introduced the effective objects which compactly combine the electromagnetic, inertial and gravitational fields acting on the test particle:
\begin{eqnarray}\label{Eeff}
\bm{\mathfrak{E}}{}_{\rm eff} &=& \bm{\mathfrak{E}} + {\frac {\gamma m} e}\,\bm{g},\\
\bm{\mathfrak{B}}{}_{\rm eff} &=& \bm{\mathfrak{B}} + {\frac {\gamma m} e}\left(
\bm{\omega} + {\frac {\bm{g}\times\widehat{\bm{v}}}{c^2}}\right).\label{Beff}
\end{eqnarray}
The effects of the magnetic and electric dipole moments are encoded in another effective object
\begin{equation}\label{Delta}
\bm{\Delta} := \bm{\mathcal M} + {\frac 1c}\,\widehat{\bm{v}}\times\bm{\mathcal P}.
\end{equation}

The physical components of spin are not $S^\alpha$, but the ``internal angular momentum" which is determined with respect to the rest frame of a particle. We denote this physical spin by $s^\alpha$. The transition to the rest frame is realized with the help of the Lorentz transformation $u^\alpha = \Lambda^\alpha{}_\beta {\stackrel \circ u}{}^\beta$ where
\begin{equation}\label{Lambda}
\Lambda^\alpha{}_\beta = \left(\begin{array}{c|c}\gamma & \gamma \widehat{v}_b/c^2 \\
\hline \gamma \widehat{v}^a & \delta^a_b 
+ (\gamma - 1)\widehat{v}^a\widehat{v}_b/\widehat{v}^2\end{array}\right).
\end{equation}
One can check that this transformation yields ${\stackrel \circ u}{}^\alpha  =
\{ 1, \bm{0}\}$, i.e., the particle is indeed at rest.

The components of the physical spin $s^\alpha$ are then obtained from $S^\alpha = \Lambda^\alpha{}_\beta s^\beta$. It is straightforward to see that $s^\alpha = \{0, \bm{s}\}$, and dynamics of the physical spin is described by the equation
\begin{equation}
{\frac {d\bm{s}}{d\tau}} = \gamma\bm{\Omega}\times\bm{s}.\label{dsp}
\end{equation}
From Eqs. (\ref{dg0})-(\ref{dgsa}) with the help of a direct (but rather lengthy) computation
we find the precession velocity
\begin{eqnarray}
\bm{\Omega} &=& {\frac em}\left\{-\,{\frac 1\gamma}\,\bm{\mathfrak{B}}{}_{\rm eff}
+ {\frac {1}{\gamma + 1}}{\frac {\widehat{\bm{v}}\times\bm{\mathfrak{E}}{}_{\rm eff}}{c^2}}
\right\}\nonumber\\
&& -\,{\frac {2}{\hbar}}\left\{\bm{\Delta} - {\frac {\gamma}{\gamma + 1}}
\,\widehat{\bm{v}}\,{\frac {\bm{\Delta}\cdot\widehat{\bm{v}}}{c^2}}\right\}.\label{OM}
\end{eqnarray}
The first term encompasses all possible (electromagnetic, inertial and gravitational) effects on particle's spin, and the second term accounts for the magnetic and electric dipole effects.

Substituting (\ref{Eeff}), (\ref{Beff}) and (\ref{Delta}), we obtain explicitly
\begin{eqnarray}
\bm{\Omega} &=& {\frac em}\left\{-\,{\frac 1\gamma}\,\bm{\mathfrak{B}}
+ {\frac {1}{\gamma + 1}}{\frac {\widehat{\bm{v}}\times\bm{\mathfrak{E}}}{c^2}}\right\}
- \bm{\omega} + {\frac {2\gamma + 1}{\gamma + 1}}\,{\frac {\widehat{\bm{v}}\times\bm{g}}{c^2}}
\nonumber\\
&& -\,{\frac {2\mu'}{\hbar}}\left\{\bm{\mathfrak{B}} - {\frac {\widehat{\bm{v}}
\times\bm{\mathfrak{E}}}{c^2}} - {\frac {\gamma}{\gamma + 1}}\,\widehat{\bm{v}}
\,{\frac {\bm{\mathfrak{B}}\cdot\widehat{\bm{v}}}{c^2}}\right\}\nonumber\\
&& -\,{\frac {2\delta'}{\hbar}}\left\{\bm{\mathfrak{E}} + \widehat{\bm{v}}
\times\bm{\mathfrak{B}} - {\frac {\gamma}{\gamma + 1}}\,\widehat{\bm{v}}
\,{\frac {\bm{\mathfrak{E}}\cdot\widehat{\bm{v}}}{c^2}}\right\}.\label{OMex}
\end{eqnarray}
Alternatively, one can use slightly different formulas by replacing $\mu'$ and $\delta'$ with the help of the definitions (\ref{mude}). Taking into account the relation between the proper and the coordinate time, ${\frac {d}{d\tau}} = \gamma {\frac {d}{dt}}$, the precession equation (\ref{dsp}) is recast into
\begin{equation}
{\frac {d\bm{s}}{dt}} = \bm{\Omega}\times\bm{s}.\label{dspt}
\end{equation}

It is worthwhile to recall that $\widehat{\bm{v}}$ , $\bm{\mathfrak{E}}$, and $\bm{\mathfrak{B}}$ are all defined with respect to the anholonomic Lorentz frame. They are related to the fields and velocity in the coordinate frame by means of (\ref{vhatLT}) and (\ref{EB1}). For the Earth, with very good accuracy one can put $V = W = 1$.

The spin dynamics (\ref{OMex}), (\ref{dspt}) is consistent in the classical and quantum pictures. For the quantum fermion particle, the Foldy-Wouthuysen Hamiltonian yields the same spin precession angular velocity when the quantum operators are replaced with the classical observables in the semi-classical limit.

\section{Manifestations of Earth's rotation}\label{rotman}

When particles move in accelerators and storage rings, it is more convenient to describe their motion relative detectors. In this case, one should subtract the angular velocity of the particle revolution from the $\bm\Omega$ given by Eq. (\ref{OMex}) and should use the cylindrical or Frenet-Serret coordinate systems. The description of the spin motion in the two coordinate systems differ (see Refs. \cite{RPJSTAB,JINRLettCylr} and references therein).

When the Frenet-Serret coordinates are used, we need to find the angular velocity of rotation of the unit vector
\begin{equation}
\widehat{\bm{N}} = {\frac {\bm{\pi}}{\pi}} = {\frac {\widehat{\bm{v}}}{\widehat{v}}}.\label{Nva}
\end{equation}
By construction, we have $(\widehat{\bm{N}}\cdot\widehat{\bm{N}}) = 1$. This unit vector determines the direction of the motion. Making use of (\ref{dvtLTA}), a direct computation yields
\begin{equation}
{\frac {d\widehat{\bm{N}}}{dt}} = \widehat{\bm{O}}\times\widehat{\bm{N}},\label{dNdta}
\end{equation}
where the rotation of the direction is determined by
\begin{eqnarray}
\widehat{\bm{O}} &=& {\frac {e}{m\gamma}}\left\{ -\,\bm{\mathfrak B} +
{\frac {\widehat{\bm{v}}}{\widehat{v}{}^2}}\times\bm{\mathfrak E}\right\}\nonumber\\
&& -\,\bm{\omega} + {\frac {\widehat{\bm{v}}}{\widehat{v}{}^2}}\times\bm{g}
\left(1 + {\frac {\widehat{v}{}^2}{c^2}}\right).\label{Oa}
\end{eqnarray}

The dynamics of spin in the Frenet-Serret coordinates is determined by the angular velocity
\begin{eqnarray}
\bm{\Omega}^{FS} &=& \bm{\Omega} - \widehat{\bm{O}} \nonumber\\
&=& -\,{\frac 1{\gamma^2 - 1}}\,{\frac {\widehat{\bm{v}}}{c^2}}\times
\left\{\,{\frac em}\bm{\mathfrak{E}} + \gamma\bm{g}\right\}\nonumber\\
&& -\,{\frac {2\mu'}{\hbar}}\left\{\bm{\mathfrak{B}} - {\frac {\widehat{\bm{v}}
\times\bm{\mathfrak{E}}}{c^2}} - {\frac {\gamma}{\gamma + 1}}\,\widehat{\bm{v}}
\,{\frac {\bm{\mathfrak{B}}\cdot\widehat{\bm{v}}}{c^2}}\right\}\nonumber\\
&& -\,{\frac {2\delta'}{\hbar}}\left\{\bm{\mathfrak{E}} + \widehat{\bm{v}}
\times\bm{\mathfrak{B}} - {\frac {\gamma}{\gamma + 1}}\,\widehat{\bm{v}}
\,{\frac {\bm{\mathfrak{E}}\cdot\widehat{\bm{v}}}{c^2}}\right\}.\label{OMFS}
\end{eqnarray}
Making use of (\ref{mude}), we have, equivalently,
\begin{eqnarray}
\bm{\Omega}^{FS} &=& -\, \frac{\gamma}{\gamma^2-1}
{\frac {\widehat{\bm{v}}\times\bm{g}}{c^2}}+{\frac {e}{m}}\left(a-\frac{1}{\gamma^2-1}
\right){\frac {\widehat{\bm{v}} \times\bm{\mathfrak{E}}}{c^2}} \nonumber\\
&& -\,a\, {\frac {e}{m}}\left(\bm{\mathfrak{B}} - {\frac {\gamma}{\gamma + 1}}
\,\widehat{\bm{v}}\,{\frac {\bm{\mathfrak{B}}\cdot\widehat{\bm{v}}}{c^2}}\right)\nonumber\\
&& -\,b\,{\frac {e}{mc}}\left\{\bm{\mathfrak{E}} + \widehat{\bm{v}}
\times\bm{\mathfrak{B}} - {\frac {\gamma}{\gamma + 1}}\,\widehat{\bm{v}}
\,{\frac {\bm{\mathfrak{E}}\cdot\widehat{\bm{v}}}{c^2}}\right\}.\label{OMFSblue}
\end{eqnarray}
Equations (\ref{OMex}), (\ref{OMFS}), and (\ref{OMFSblue}) are compatible with corresponding equations in classical electrodynamics (see Refs. \cite{T-BMT3,FukuyamaSilenko,PhysScr} and references therein).

It is worthwhile to note that Eqs. (\ref{OMFS}), (\ref{OMFSblue}) can be compactly re-written
in terms of the effective objects:
\begin{eqnarray}
\bm{\Omega}^{FS} &=& -\,{\frac {1}{\gamma^2 - 1}}\,{\frac em}
\,{\frac {\widehat{\bm{v}}\times\bm{\mathfrak{E}}{}_{\rm eff}}{c^2}}\nonumber\\
&& - {\frac {2}{\hbar}}\left\{\bm{\Delta} - {\frac {\gamma}{\gamma + 1}}
\,\widehat{\bm{v}}\,{\frac {\bm{\Delta}\cdot\widehat{\bm{v}}}{c^2}}\right\}.\label{OFS}
\end{eqnarray}

Quite remarkably, the explicit contribution of the Earth's rotation $-\,\bm{\omega}$ disappeared from $\bm{\Omega}^{FS}$. Nevertheless, the effects of rotation are still present implicitly in the expressions $\widehat{\bm{v}} = \bm{v} + \bm{\omega}\times\bm{r}$ and in $\bm{\mathfrak{E}} = \bm{E} - (\bm{\omega}\times\bm{r})\times\bm{B}$ (however, notice that $\bm{\mathfrak{B}} = \bm{B}$).

One can easily understand why the angular velocity of the spin motion relative to the beam trajectory, $\bm{\Omega}^{FS}$, is not equal to $\bm{\Omega} - \bm{O}$, where $\bm{O}$ is defined by
\begin{equation}
{\frac {d\bm{N}}{dt}} = \bm{O}\times\bm{N}, \qquad \bm{N} = {\frac {\bm{v}}{v}}. \label{dNdtbeta}
\end{equation}

Let us consider the frozen spin ring \cite{EDM,OMS,Anastassopoulos} in the Minkowski frame. In this ring, the angle between the spin and the tangent line to the particle trajectory is constant. Therefore, $\bm{\Omega}^{FS}=0$. We can now proceed to the frame which rotates about the vertical axis passing through the center of the ring. In this frame, the angle between the spin and the tangent line to the particle trajectory remains constant.

The velocity and spin evolution in the rotating frame is defined by
\begin{equation}\label{dNdvspin}
{\frac {d\bm{v}}{dt}} = -\bm{2\omega}\times\bm{v}, \quad {\frac {d\widehat{\bm{v}}}{dt}}
= -\bm{\omega}\times\widehat{\bm{v}}, \quad {\frac {d\bm{s}}{dt}} =  -\bm{\omega}\times\bm{s}.
\end{equation}

Equations (\ref{dNdvspin}) show that the use of the definition $\bm{\Omega}^{FS} = \bm{\Omega} - \widehat{\bm{O}}$ yields the correct value $\bm{\Omega}^{FS}=0$, while the quantity $\bm{\Omega} - \bm{O}$ is equal to $\bm{\omega}$ and is not applicable to the description of the spin motion relative to the beam trajectory. We note the inaccuracy in Ref. \cite{PRD2007}.

The rotation of the Earth manifested in some experiments with particles and beams \cite{Werner79,sil1,sil2,sil3}. For the first time, the manifestation of Earth's rotation in such experiments has been discovered by Werner, Staudenmann, and Colella \cite{Werner79} by means of the neutron interferometry. In neutron interference experiments, the period of particle revolution on a closed path depends on a rotation direction (clockwise or counterclockwise) due to the Sagnac effect \cite{Sagnac} (see also Ref. \cite{Sagnacm}). This effect is caused by the rotation of the lab relative to the ``immobile'' far stars. The theory of the Sagnac effect is presented in details in Refs. \cite{Sagnacm,AHL,RauchWerner}. While the \emph{world} velocity of the particle depends on the Earth's rotation, the \emph{measurable} velocity cannot exceed $c$ (see explanations given in Refs. \cite{SagnacLogunov,SagnacMalykin,Sagnacl}). Some problems connected with the Sagnac effect have also considered in Refs. \cite{Sagnacn,Sagnack}. The review of experiments performed is given in Ref. \cite{Werner08}.

One can usually neglect the effects of rotation in storage ring experiments. In particular, they are negligible in all \emph{g}-2 experiments. These effects can be of minor importance in storage-ring EDM experiments based on the frozen spin method. However, the contribution of the Earth rotation is very important in EDM experiments with atoms at rest \cite{Venema} and in some other experiments with atoms \cite{Gemmel,Gemmel2,Gemmel3,Gemmel4,Gemmel5}. The contribution of the Earth rotation was taken into account to obtain the restrictions on the hypothetical interactions. In particular, the analysis of the experimental data \cite{Gemmel,Gemmel2,Gemmel3,Gemmel4,Gemmel5} produced the new bounds on the spacetime torsion \cite{ostor}. It is important that Cartan's torsion of spacetime couples only to particle's spin but not to the orbital angular momentum of a test particle \cite{HehlObukhovPuetzfeld}. If the spacetime torsion is nonzero, the angular velocity of the atom spin precession in the Earth's rotating frame is different from $-\bm\omega$.

As a rule, the vectors $\bm\Omega$ and $\bm\omega$ are not collinear. When the vector $\bm\Omega$ is vertically directed (this always takes place in storage rings), only the vertical component of the angular velocity of the Earth rotation and the radial component of the Coriolis acceleration are important. The contribution of the Earth rotation to the angular frequency of the spin precession is equal to $\omega\sin{\varphi}$, where $\varphi$ is the geographic latitude. For atoms at rest, the horizontal component of $\bm\omega$ can imitate the presence of the EDM. For particles and nuclei in storage rings, this component may be disregarded because the radial and longitudinal projections of the angular velocity of the Earth rotation, $\omega_r$ and $\omega_\phi$, are zero on the average.

\section{Manifestations of Earth's gravity}\label{gravman0}

Manifestations of the Earth gravity in storage rings and accelerators have been analyzed in Ref. \cite{PRD2007}. However, this analysis does not take into account an influence of gravity on electromagnetic interactions. It has been shown in Sec. \ref{rotatframe} that a non-Minkowskian metric can significantly change these interactions. To evaluate the influence of gravity on electromagnetic interactions, we can use the result obtained in Ref. \cite{ostrong2}. In the squared Dirac equation, spin effects in electromagnetic interactions are defined by spin-electromagnetic field coupling of the form $\sigma^{\alpha\beta}F_{\alpha\beta}$, where $F_{\alpha\beta}=e_\alpha^i e_\beta^j F_{ij}$ is the electromagnetic field tensor convoluted with a tetrad and the spin operator $\sigma^{\alpha\beta}=\frac i2(\gamma^\alpha\gamma^\beta - \gamma^\beta\gamma^\alpha)$ is constructed from the usual Dirac matrices \cite{ostrong2} of the flat Minkowski space. In the non-inertial reference frame in the presence of the gravitational field, $F_{\alpha\beta}$ differs from Maxwell's electromagnetic field tensor $F_{ij}$. The difference is encoded in the tetrad fields.

The inertial and gravitational fields affect the electromagnetic fields as a continuous medium with nontrivial polarizability and magnetizability properties. The Maxwell equations have the usual form
\begin{eqnarray}
\bm{\nabla}\times \bm{E} + \dot{\bm{B}} = 0,
\qquad \bm{\nabla}\cdot\bm{B} = 0,\label{maxFR}\\
\bm{\nabla}\times \bm{H} - \dot{\bm{D}} = \bm{J},
\qquad \bm{\nabla}\cdot\bm{D} = \rho,\label{maxHR}
\end{eqnarray}
where nabla $\bm{\nabla} = \{\partial_a\}$, the dot denotes the time derivative, $\dot{\ } = \partial/\partial t$, and $\bm{J}, \rho$ are the current and charge densities of the sources. The inertial and gravitational influence is encoded in the constitutive relations between the electric and magnetic fields $\bm{E}, \bm{B}$ and the electric and magnetic excitations $\bm{D}, \bm{H}$. For the metric of the Earth (\ref{ds2}), with (\ref{VLT}) and (\ref{WLT}), the constitutive relations read explicitly:
\begin{eqnarray}
\bm{D} &=& \varepsilon_0{\frac WV}\left[\bm{E} -
(\bm{\omega}\times\bm{r})\times\bm{B}\right],\label{constR1}\\
\bm{H} &=& {\frac 1{\mu_0}}{\frac VW}\bm{B}\nonumber\\
&& -\,\varepsilon_0{\frac WV} (\bm{\omega}\times\bm{r})\times\left[\bm{E} -
(\bm{\omega}\times\bm{r})\times\bm{B}\right].\label{constR2}
\end{eqnarray}
Here $\varepsilon_0$ and $\mu_0$ are the electric and magnetic constants of vacuum (not to confuse the latter with the magnetic dipole moment!).

One can analyze independently the rotational and gravitational contributions to the dynamics of particle's momentum and spin. The gravitational corrections to Maxwell's electrodynamics are extremely small, see (\ref{GMC2R}), and with a good accuracy we can put $V = W = 1$ for the Earth.

The effects of rotation were discussed in the previous section, and now we specialize to the purely gravitational effects. For the Schwarzschild metric in the isotropic coordinates, Eq. (\ref{ds2}) takes the form $ds^2 = V^2c^2dt^2 - W^{2}(dx^a)^2$, with (\ref{VLT}) and (\ref{WLT}). In this case, (\ref{vhatLT}) and (\ref{EB1}) reduce to
\begin{equation}
\bm{\mathfrak{E}}=\frac{\bm{E}}{VW},\quad \bm{\mathfrak{B}}=\frac{\bm{B}}{W^2},
\quad  \widehat{\bm{v}}=\frac{W}{V}\bm{v}.\label{dsthree}
\end{equation}
The contribution of electromagnetic interactions to the particle motion is defined by
\begin{equation}\label{clfLL}
F^{ak}u_k = g^{ai}F_{ik}u^k = -\,{\frac{u^0}{W^2}}\left(\bm{E} + \bm{v}\times\bm{B}\right)^a.
\end{equation}

It is easy to see that in the case under consideration $\bm N=\widehat{\bm N}$ and  $\bm O=\widehat{\bm O}$. It can be checked that
\begin{eqnarray}
{\frac{1}{W^2}}\left({\bm B} - \frac{\bm v\times{\bm E}}{v^2}\right) = \bm{\mathfrak{B}}
- \frac{\hat{\bm v}\times\bm{\mathfrak{E}}}{\hat{v}^2}.\label{eqO3}
\end{eqnarray}

The particle motion is defined by Eq. (\ref{Oa}). We underline that the vector $\hat{\bm v}\times\bm{g}$ is directed radially.

For particles in accelerators and storage rings, the Newton-like gravitational force (a relativistic generalization of Newton's force) in the weak-field approximation is given by \cite{PRD2007,PRD}
\begin{equation}
\bm F_g =\frac{2\gamma^2-1}{\gamma} m\bm{g}.\label{Fg0}
\end{equation}

The gravitational force causes a vertical shift of the particle orbit and is counterbalanced by a force created by the focusing system. When the magnetic focusing is used, the latter force is the Lorentz one. Since $\bm F_m=-\bm F_g$, the average radial magnetic field reads
\begin{eqnarray}
B_r=\frac{(2\gamma^2-1)mg}{c\sqrt{\gamma^2-1}|e|}.\label{fm0}
\end{eqnarray}
This field determines the spin rotation with the average angular velocity \cite{PRD2007}
\begin{eqnarray}\label{eql0}
\bm\Omega_{m}=-\,\frac{(1+a\gamma)(2\gamma^2-1)}{c^2(\gamma^2-1)}\hat{\bm v}\times\bm{g}.
\end{eqnarray}

When the electric focusing is used, the gravitational force (\ref{Fg0}) gives rise to the vertical electric field
\begin{eqnarray}
E_z=\frac{(2\gamma^2-1)mg}{\gamma e}.\label{fe0}
\end{eqnarray}
The corresponding average angular velocity of the spin rotation is given by
\begin{eqnarray}
\bm\Omega_{e}=-\,\frac{2\gamma^2-1}{c^2\gamma}\left(a + \frac{1}{\gamma+1}\right)
\hat{\bm{v}}\times\bm{g}. \label{eq20}
\end{eqnarray}

The total effect of the Earth gravity on the spin motion reads
\begin{eqnarray}
\bm\Omega_g + \bm\Omega_{m} &=& -\,\frac{\gamma\left[1+a(2\gamma^2-1)\right]}{c^2(\gamma^2-1)}
\hat{\bm{v}}\times\bm{g},\label{OmegM0}\\
\bm\Omega_g + \bm\Omega_{e} &=& \frac{1-a(2\gamma^2-1)}{c^2\gamma}\hat{\bm{v}}
\times\bm{g}.\label{Omeg0}
\end{eqnarray}

Thus, the total contribution of the Earth gravity is equal to $$\bm\Omega'_g= \bm\Omega_g+\bm\Omega_{m}$$ and $$\bm\Omega'_g=\bm\Omega_g+\bm\Omega_{e}$$ when one uses the magnetic and the electric focusing, respectively.   One should add $\bm\Omega_{e}$ or $\bm\Omega_{m}$ to the electromagnetic part of Eqs. (\ref{OMex}), (\ref{OMFS})-(\ref{OFS}) or should substitute $\bm\Omega'_g$ for $\bm\Omega_g$ into this equation. The additional torques governing the spin are caused by the corresponding focusing field and by the geodetic effect (the spin precession in a gravitational field). These torques leads to the spin rotation about the radial axis \cite{PRD} which angular velocity is given by Eqs. (\ref{OmegM0}) and (\ref{Omeg0}).

It is instructive to mention that the change of the constitutive tensor of Maxwellian electrodynamics due to the Earth rotation is $\omega c/g\approx2200$ times greater than the corresponding change due to the Earth gravity.

\section{Corrections for the rotation and gravity of the Earth in high-energy physics experiments}\label{spineqn}

Let us consider effects of the rotation and gravity of the Earth in high-energy physics experiments. Evidently, the resulting equations of motion differ from corresponding equations in Minkowski frames. To determine effects of the rotation and the gravity, we need to specify measured quantities.

As a rule, the magnetic field in high-precision experiments is measured with magnetometers based on the spin precession of nuclei at rest (see, e.g., Refs. \cite{Venema,Gemmel,Gemmel2,Gemmel3,Gemmel4,Gemmel5,PRDfinal}). When the electric field is equal to zero ($\bm E=0$), we obtain $\bm{v}=0$ and $\bm{\mathfrak{E}}=- (\bm\omega\times\bm{r})\times\bm{B}$.  If the magnetic field is orthogonal to $\bm\omega\times\bm{r}$, the spin precession in a magnetometer is given by
\begin{equation}\label{OMagn}
\bm{\Omega} = -\,\left[1 - {\frac {(\bm\omega\times\bm{r})^2}{c^2}}\right]
{\frac {e(1 + a)\bm{\mathfrak{B}}}{m}}.
\end{equation}
In this equation, the inertial and gravitational contributions which can be taken into account as systematical corrections are omitted and the EDM terms are neglected.

Thus, the correction to the measured magnetic field caused by Earth's rotation is nonzero while it is negligible in contemporary high-energy physics, since $(\bm\omega\times\bm{r})^2/c^2\sim10^{-12}$. In particular, the precision in the past and planned muon $g$-2 experiments is less than $1\times10^{-7}$. The correction disappears in experiments with atoms and nuclei at rest when the magnetometers based on the spin precession are used.

A determination of the electric field is based on a measurement of a force acting on a charge at rest ($\bm v=0$). If $\bm{\mathfrak{B}}=\bm{B}=0$, the determined electric field does not substantially differ from $\bm E$ [see Eq. (\ref{EduFa})]. This equation shows the corrections caused by the Newton-like gravitational force, by the inertial centrifugal and Coriolis forces, and by the non-Maxwellian effects in electrodynamics. The latter effects are described by the last line in Eq. (\ref{EduFa}) and are given by the term $u^0e(\bm\omega\times\bm{r})\bigl(\bm E\cdot(\bm\omega\times\bm{r})\bigr)\bigr/(mc^2)$. This term is very small and defines a correction of the order of $10^{-12}$ to the Coulomb force.

We can now discuss effects of terrestrial rotation and gravity in the high-energy physics experiments. As an example, let us consider the \emph{g}-2 and EDM experiments performed in storage rings. In the muon EDM experiment fulfilled in the Brookhaven National Laboratory (BNL) \cite{PRDfinal}, the Lorentz factor of muons rotating in a uniform magnetic field satisfied the condition
\begin{eqnarray}
 a = {\frac {2\mu' m}{e\hbar}}=\frac{1}{\gamma^2-1}. \label{Lentf0}
\end{eqnarray}
Under this condition, the spin precession angular velocity (\ref{OMFSblue}) reduces to
\begin{eqnarray}
\bm{\Omega}^{FS} &=& -\,a\left\{\gamma
{\frac {\widehat{\bm{v}}\times\bm{g}}{c^2}}
+ {\frac {e}{m}}\left(\bm{\mathfrak{B}} - {\frac {\gamma}{\gamma + 1}}\,\widehat{\bm{v}}
\,{\frac {\bm{\mathfrak{B}}\cdot\widehat{\bm{v}}}{c^2}}\right)\right\}\nonumber\\
&& -\,b\,{\frac {e}{mc}}\left\{\bm{\mathfrak{E}} + \widehat{\bm{v}}
\times\bm{\mathfrak{B}} - {\frac {\gamma}{\gamma + 1}}\,\widehat{\bm{v}}
\,{\frac {\bm{\mathfrak{E}}\cdot\widehat{\bm{v}}}{c^2}}\right\}.\label{OMFS1}
\end{eqnarray}

Equation (\ref{OMFS1}) together with the equations of motion (\ref{EduF0}) and (\ref{EduFa}) shows that the local corrections to the \emph{g}-2 frequency are of the order of $|\bm\omega\times\bm{r}|/c\sim10^{-6}$.  Corrections of such an order are not too small and should be taken into consideration.

A contribution of Earth's rotation to dynamics of momentum and spin is defined by nondiagonal components of the metric taken in any fixed point of a lab, $\bm{r}_0$, and by a change of these components inside of the lab. When one considers a particle in an accelerator or a storage ring, we have $\bm{r}=\bm{r}_0+\bm{\mathfrak{r}}$, where $\bm{\mathfrak{r}}$ is the radius vector of the particle relative to the chosen point of the lab. 

A simple analysis demonstrates that terms proportional to $\bm{r}_0$ average to zero because the average value of $\bm{v}$ is zero. Terms proportional to $\bm{\mathfrak{r}}$ may be nonzero after averaging but they are usually negligible. For $\mathfrak{r}\sim10$~m, we find $|\bm\omega\times\bm{\mathfrak{r}}|/c\sim10^{-12}$. Perhaps it is reasonable to take into account the above-mentioned terms in some EDM and other experiments in storage rings. In larger labs and rings like LHC the effects are enhanced accordingly. 

In the planned J-PARC muon {\emph g}-2 experiment \cite{JPARC}, the condition (\ref{Lentf0}) is not satisfied ($p=300$ MeV/$c$). Nevertheless, all corrections for the rotation and gravity of the Earth can be neglected.

The cancellation of gravitational corrections is a nontrivial effect. Nominal order of magnitude of the gravitational corrections is $r_g/r\sim 10^{-9}$. Equations (\ref{dsthree})-(\ref{eqO3}) show that the corrections to the electromagnetic interactions caused by Earth's gravity are of the same order of magnitude. The electromagnetic fields governing the momentum and the spin are, in principle, different. If the difference between them were not canceled in the expression for $\bm \Omega-\widehat{\bm{O}}$, the gravitational corrections would be of the orders of $r_g/r$ and $r_g/(ra\gamma)$ for the angular frequency of the spin precession in the Cartesian coordinates and for the {\emph g}-2 frequency, respectively. The latter quantity would be of the order of $0.01\div0.1$ ppm for the BNL muon {\emph g}-2 experiment and of the order of $0.1\div1$ ppm for the J-PARC one. However, the gravitational corrections are canceled.

A part of muon {\emph g}-2 experiments is a search for the muon EDM \cite{MuEDM08}. In principle, the gravitational contribution to the spin precession defined by Eqs. (\ref{OmegM0}) and (\ref{Omeg0}) can imitate the presence of the EDM. However, the corresponding false EDM is very small. In the BNL muon {\emph g}-2 experiment, it is of the order of $10^{-30}$ $e\cdot$cm while the upper limit \cite{MuEDM08} on the muon EDM is $|\delta'_\mu|<1.9\times10^{-19}$ $e\cdot$cm. Nevertheless, this effect may not be negligible in specially designed EDM experiments performed with the frozen spin method \cite{EDM}. A calculation shows that the Earth gravity can produce the same effect as deuteron's EDM of $\delta' = 1.5\times10^{-29}$ $e\cdot$cm in the planned dEDM experiment with magnetic focusing \cite{OMS}. The effect of Earth's gravity manifests in the additional spin rotation about the radial axis and can be important, because the expected sensitivity of this experiment \cite{OMS} is of the same order. The influence of Earth's gravity on the angular velocity of the spin precession in \emph{g}-2 experiments can be neglected.

The influence of Earth's rotation on spin dynamics can be more important in storage ring EDM experiments than in {\emph g}-2 ones. The use of the frozen spin method makes slow rotations of the spin about the vertical axis more significant. These rotations are due to the corrections to the constitutive law in the Maxwell-Lorentz electrodynamics.

Terms proportional to $\bm{\mathfrak{r}}$ can contribute to the Sagnac effect. This effect may be observable for particles and nuclei in storage rings. While particles and nuclei pass two halves of the orbit during different intervals of time, the main correction proportional to the radius of the Earth vanishes. However, a smaller correction proportional to the radius of the ring is nonzero and can be important in the EDM experiments \cite{Melissinos,Semertzidis}. Evidently, the above-mentioned terms change the contribution of Earth's rotation to the angular velocities of the beam revolution for clockwise and counterclockwise beams. The angular frequencies of the spin rotation and the beam revolution are not conserved when the directions of the magnetic field and the beam velocity are reversed. If the opposite directions of rotation are due to the different charges of particles (say, $e^+$ and $e^-$), this would simulate \cite{PRD2007} the \emph{CP}-violation caused by Earth's rotation.

\section{Conclusion}\label{conc}

We analyzed the possible effects of the rotation and gravitational field of the Earth on the particle motion and the spin evolution. This influence is not negligible and should be taken into account in high-energy physics experiments. We studied the manifestations of the inertial and gravitational effects in the framework of the general-relativistic Dirac theory for a fermion particle with spin 1/2. In astrophysical context, a similar analysis was performed for the dynamics of the massive neutrinos in a non-inertial frame of a rotating background matter \cite{Dvor1,Dvor2} and in anisotropic universes \cite{KamTeryaev1,KamTeryaev2}.

We have derived for the first time the relativistic Foldy-Wouthuysen Hamiltonian for the Dirac particle with the AMM and EDM in the rotating frame and have determined the electromagnetic fields governing the dynamics of the momentum and the spin of the particle. The results obtained have been used for the analysis of systematic corrections due to the terrestrial rotation and gravity in the {\emph g}-2 and EDM experiments. In particular, for the first time we have obtained the equations for the spin precession when the inertial, gravitational and electromagnetic fields are present simultaneously. All the results have been obtained by both classical and quantum mechanical methods and their complete agreement has been found.

Earth's rotation gives rise to the Coriolis and the centrifugal forces in the lab frame and also manifests itself in the additional rotation of the spin and in the change of the constitutive tensor of Maxwell's electrodynamics. The corrections to the motion of the particle and the spin due to Earth's rotation are rather small. Bigger corrections are oscillatory and their contributions average to zero. Other corrections due to the inhomogeneity of the inertial field are not oscillatory but they are very small and their consideration may be reasonable only for the storage ring EDM experiments. 

 Earth's gravity manifests in the additional force acting on the particle momentum, in the additional torque acting on the spin, and in the change of the constitutive tensor of Maxwell's electrodynamics. However, so far there are no observable indications of electromagnetic effects caused by Earth's gravity. The additional forces are the Newton-like force and the reaction force provided by a focusing system. The additional torques are caused by the focusing field and by the geodetic effect. In the storage ring EDM experiments, these forces and torques lead to the additional spin rotation about the radial axis.

\section*{Acknowledgments}

A.J.S. is grateful to Y. K. Semerzidis for valuable comments and discussions and acknowledges the hospitality and support of the Center for Axion and Precision Physics Research at the Korea Advanced Institute of Science and Technology. Y.N.O. thanks Bahram Mashhoon for the helpful discussions and for the useful literature hints. The work was supported in part by the Belarusian Republican Foundation for Fundamental Research (Grant No. $\Phi$16D-004), by the Heisenberg-Landau program of the German Ministry for Science and Technology (Bundesministerium f\"{u}r Bildung und Forschung), by PIER (``Partnership for Innovation, Education and Research'' between DESY and Universit\"at Hamburg), and by the Russian Foundation for Basic Research (Grant No. 14-01-00647 and Grant No. 16-02-00844-A).


\begin{thebibliography}{99}

\bibitem{Sagnac}
G. Sagnac, L'\'ether lumineux d\'emontr\'e par l'effet du vent
relatif l'\'ether dans un interf\'erometre en rotation uniforme,
Compt. Rend. Acad. Sci. (Paris) \textbf{157}, 708 (1913).

\bibitem{ostrong}
Yu.N. Obukhov, A.J. Silenko, and O.V. Teryaev,
Spin dynamics in gravitational fields of rotating bodies and the
equivalence principle, Phys. Rev. D \textbf{80}, 064044 (2009).

\bibitem{ostrong2}
Yu.N. Obukhov, A.J. Silenko, and O.V. Teryaev,
Dirac fermions in strong gravitational fields,
Phys. Rev. D \textbf{84}, 024025 (2011).

\bibitem{ostgrav}
Yu.N. Obukhov, A.J. Silenko, and O.V. Teryaev,
Spin in an arbitrary gravitational field,
Phys. Rev. D {\bf 88}, 084014 (2013).

\bibitem{ostor}
Yu.N. Obukhov, A.J. Silenko, and O.V. Teryaev,
Spin-torsion coupling and gravitational moments of Dirac fermions:
Theory and experimental bounds, Phys. Rev. D \textbf{90}, 124068 (2014).

\bibitem{ostor2}
Yu.N. Obukhov, A.J. Silenko, and O.V. Teryaev,
Spin-gravity interactions and equivalence principle,
Int. J. Mod. Phys.: Conf. Ser. \textbf{40}, 1660081 (2016).

\bibitem{HehlObukhov}
F. W. Hehl and Yu. N. Obukhov, \emph{Foundations of Classical Electrodynamics:
Charge, Flux, and Metric}, (Birkh\"{a}user: Boston, 2003). 410 p.

\bibitem{Schmutzer1}
E. Schmutzer,
Maxwell theory and quantum mechanics in a rotating frame of reference,
in: \emph{``Symposia mathematica''} (Academic Press: London, 1973) vol. 12, pp. 281-296.

\bibitem{Schmutzer2}
E. Schmutzer, Maxwell-Theorie (in Medien) und Quantentheorie in einem
rotierenden Bezugssystem, Ann. Phys. (Leipzig) {\bf 29}, 75-95 (1973).

\bibitem{Schmutzer3}
E. Schmutzer and J. Pleba\'nski,
Quantum mechanics in non-inertial frames of reference,
Fortschr. Physik {\bf 25}, 37-82 (1977).

\bibitem{Quantmf}
G. Papini, Quantum physics in inertial and gravitational fields, in:
\emph{Relativity in Rotating Frames: Relativistic Physics in Rotating Reference Frames},
G. Rizzi, M. L. Ruggiero, eds. (Kluwer Acad. Publ.: Dordrecht, 2004), pp. 335-360.

\bibitem{Quantmn}
J. Anandan and J. Suzuki, Quantum mechanics in a rotating frame, in:
\emph{Relativity in Rotating Frames: Relativistic Physics in Rotating Reference Frames},
G. Rizzi, M. L. Ruggiero, eds. (Kluwer Acad. Publ.: Dordrecht, 2004), pp. 361-370.

\bibitem{Itzykson}
C. Itzykson and J.-B. Zuber, {\it Quantum field theory} (McGraw-Hill: New York, 1980).

\bibitem{RPJ}
A.J. Silenko, Quantum-mechanical description of the electromagnetic
interaction of relativistic particles with electric and magnetic dipole moments,
Russ. Phys. J. \textbf{48}, 788 (2005).

\bibitem{Warszawa}
A.J. Silenko, Classical and quantum spins in curved spacetimes,
Acta Phys. Polon. B Proc. Suppl. \textbf{1}, 87 (2008). 

\bibitem{FW}
L.L. Foldy and S.A. Wouthuysen,
On the Dirac theory of spin $1/2$ particles and its non-relativistic limit,
Phys. Rev. \textbf{78}, 29 (1950).

\bibitem{PRA}
A.J. Silenko, Foldy-Wouthuysen transformation and semiclassical
limit for relativistic particles in strong external fields,
Phys. Rev. A \textbf{77}, 012116 (2008).

\bibitem{FWproof}
A.J. Silenko, General method of the relativistic Foldy-Wouthuysen
transformation and proof of validity of the Foldy-Wouthuysen Hamiltonian,
Phys. Rev. A \textbf{91}, 022103 (2015).

\bibitem{NW}
T.D. Newton and E.P. Wigner, Localized states for elementary systems,
Rev. Mod. Phys. \textbf{21}, 400 (1949).

\bibitem{JMP}
A.J. Silenko, Foldy-Wouthuysen transformation for relativistic particles
in external fields, J. Math. Phys. \textbf{44}, 2952 (2003).

\bibitem{CMcK}
J.P. Costella and B.H.J. McKellar, The Foldy-Wouthuysen transformation,
Am. J. Phys. \textbf{63}, 1119 (1995).

\bibitem{JINRLett12}
A.J. Silenko, Classical Limit of Equations of the Relativistic Quantum
Mechanics in the Foldy-Wouthuysen Representation,
Phys. Part. Nucl. Lett. \textbf{10}, 91 (2013).

\bibitem{TMPFW}
A.J. Silenko, Comparative analysis of direct and ``step-by-step''
Foldy-Wouthuysen transformation methods,
Theor. Math. Phys. \textbf{176}, 987 (2013).

\bibitem{PRDspin1}
A.J. Silenko, Quantum-mechanical description of spin-1 particles with electric
dipole moments, Phys. Rev. D {\bf 87}, 073015 (2013).

\bibitem{PRD2007}
A.J. Silenko and O. V. Teryaev,
Equivalence principle and experimental tests of gravitational spin effects,
Phys. Rev. D {\bf 76}, 061101(R) (2007).

\bibitem{kuro}
J. Kuroiwa, M. Kasai, and F. Toshifumi,
A treatment of general relativistic effects in quantum interference,
Phys. Lett. A {\bf 182}, 330-334 (1993).

\bibitem{dimo}
S. Dimopoulos, P.W. Graham, J.M. Hogan, and M.A. Kasevich,
General relativistic effects in atom interferometry,
Phys. Rev. D {\bf 78}, 042003 (2008).

\bibitem{bahlnp}
B. Mashhoon, Quantum theory in accelerated frames of reference,
Lect. Notes Phys. {\bf 702} 112-132 (2006).

\bibitem{bahyo}
B. Mashhoon and Yu.N. Obukhov,
Spin precession in inertial and gravitational fields,
Phys. Rev. D {\bf 88}, 064037 (2013).

\bibitem{Schiff}
L.I. Schiff, A question in general relativity,
Proc. Nat. Acad. Sci. {\bf 25}, 391-395 (1939).

\bibitem{Heer}
C.V. Heer, Resonant frequencies of an electromagnetic cavity in an
accelerated system of reference. Phys. Rev. {\bf 134}, A799-A804 (1964).

\bibitem{Webster}
D.L. Webster and R.C. Witten, Which electromagnetic equations apply in
rotating coordinates? Astrophys. Space Sci. {\bf 24} 323-333 (1973).

\bibitem{Menegozzi}
L.N. Menegozzi and W.E. Lamb, Theory of ring laser,
Phys. Rev. A {\bf 8}, 2103-2125 (1973).

\bibitem{Gron}
{\O}. Gr{\o}n, Application of Schiff's rotating-frame electrodynamics,
Int. J. Theor. Phys. {\bf 23} 441-448 (1984).

\bibitem{Scorgie}
G.C. Scorgie, Electromagnetism in non-inertial coordinates,
J. Phys. A: Math. Gen. {\bf 23} 5169-5184 (1990).

\bibitem{Ciu}
I. Ciufolini, Measurement of the Lense-Thirring drag on high-altitude,
laser-ranged artificial satellites, Phys. Rev. Lett. {\bf 56} 278-281 (1986).

\bibitem{RPJSTAB}
A. J. Silenko, Equation of spin motion in storage rings in the cylindrical
coordinate system, Phys. Rev. ST Accel. Beams \textbf{9}, 034003 (2006).

\bibitem{JINRLettCylr}
A. J. Silenko, Comparison of spin dynamics in the cylindrical and Frenet-Serret
coordinate systems, Phys. Part. Nucl. Lett. \textbf{12}, 8 (2015).

\bibitem{T-BMT3}
V. Bargmann, L. Michel, and V.L. Telegdi, Precession of the polarization
of particles moving in a homogeneous electromagnetic field,
Phys. Rev. Lett. \textbf{2}, 435 (1959).

\bibitem{FukuyamaSilenko}
T. Fukuyama and A. J. Silenko, Derivation of generalized
Thomas-Bargmann-Michel-Telegdi equation for a particle with electric
dipole moment, Int. J. Mod. Phys. A \textbf{28}, 1350147 (2013).

\bibitem{PhysScr}
A. J. Silenko, Spin precession of a particle with an electric dipole moment:
contributions from classical electrodynamics and from the Thomas effect,
Phys. Scripta {\bf 90}, 065303 (2015).

\bibitem{EDM}
F. J. M. Farley, K. Jungmann, J. P. Miller, W. M. Morse, Y. F.
Orlov, B. L. Roberts, Y. K. Semertzidis, A. Silenko, and E. J.
Stephenson, New method of measuring electric dipole moments
in storage rings, Phys. Rev. Lett. {\bf 93}, 052001 (2004).

\bibitem{OMS}
D. Anastassopoulos \emph{et al} (EDM Collaboration),
{\it AGS Proposal: Search for a permanent electric dipole moment
of the deuteron nucleus at the $10^{-29} \,e \cdot\,$cm Level,}
\url{https://www.bnl.gov/edm/files/pdf/deuteron_proposal_080423_final.pdf}

\bibitem{Anastassopoulos}
V. Anastassopoulos \emph{et al.} (Storage Ring EDM Collaboration),
\emph{A storage ring experiment to detect a proton electric dipole moment},
arXiv:physics.acc-ph/1502.04317.

\bibitem{Werner79}
S. A. Werner, J.-L. Staudenmann, and R. Colella,
Effect of Earth's rotation on the quantum mechanical phase of the neutron,
Phys. Rev. Lett. {\bf 42}, 1103-1106 (1979).

\bibitem{sil1}
M.P. Silverman, Effect of the Earth's rotation on the optical properties of atoms,
Phys. Lett. A {\bf 146}, 175-180 (1990).

\bibitem{sil2}
M.P. Silverman,
Measurement of hydrogen hyperfine splittings as a test of quantum mechanics in
a noninertial reference frame, Phys. Lett. A {\bf 152}, 133-136 (1991).

\bibitem{sil3}
M.P. Silverman, Rotationally induced optical activity in atoms,
EPL (Europhysics Letters) {\bf 9}, 95-99 (1989).

\bibitem{Sagnacm}
G. Rizzi, M. L. Ruggiero, The relativistic Sagnac effect: two derivations, in:
\emph{Relativity in Rotating Frames: Relativistic Physics in Rotating Reference Frames},
G. Rizzi, M. L. Ruggiero, eds. (Kluwer Acad. Publ.: Dordrecht, 2004), pp. 179-220.

\bibitem{AHL}
J. Audretsch, F.W. Hehl, C. L\"{a}mmerzahl, Matter wave interferometry and
why quantum objects are fundamental for establishing a gravitational theory,
in: \emph{Relativistic Gravity Research with Emphasis on Experiments and Observations,
Springer Lecture Notes in Physics}, J. Ehlers and G. Sch\"{a}fer, eds.
(Springer: Berlin, 1992), p. 368-407.

\bibitem{RauchWerner}
H. Rauch and S. A. Werner, Neutron interferometry: Lessons in experimental
quantum mechanics, 2nd ed. (Oxford Univ. Press: Oxford, 2015). 447 p.

\bibitem{SagnacLogunov}
A. A. Logunov and Yu. V. Chugreev,
The Sagnac effect: correct and incorrect explanations,
Sov. Phys. Usp. \textbf{31}, 861 (1988).

\bibitem{SagnacMalykin}
G. B. Malykin, The Sagnac effect: correct and incorrect explanations,
Phys.-Usp. \textbf{43}, 1229 (2000).

\bibitem{Sagnacl}
J.-F. Pascual-S\'{a}nchez, A. S. Miguel, F. Vicente, Isotropy of the velocity
of light and the Sagnac effect, in: \emph{Relativity in Rotating Frames:
Relativistic Physics in Rotating Reference Frames}, G. Rizzi, M. L. Ruggiero, eds.
(Kluwer Acad. Publ.: Dordrecht, 2004), pp. 167-178.

\bibitem{Sagnacn}
N. Ashby, The Sagnac effect in the global positioning system, in:
\emph{Relativity in Rotating Frames: Relativistic Physics in Rotating Reference Frames},
G. Rizzi, M. L. Ruggiero, eds. (Kluwer Acad. Publ.: Dordrecht, 2004), pp. 11-28.

\bibitem{Sagnack}
F. Selleri, Sagnac effect: end of the mystery, in: \emph{Relativity in Rotating Frames:
Relativistic Physics in Rotating Reference Frames}, G. Rizzi, M. L. Ruggiero, eds.
(Kluwer Acad. Publ.: Dordrecht, 2004), pp. 57-78.

\bibitem{Werner08}
S. Werner, Does a neutron know that the Earth is rotating? Gen. Rel. Grav.
\textbf{40}, 921 (2008). 

\bibitem{Venema}
B. J. Venema, P. K. Majumder, S. K. Lamoreaux, B. R. Heckel, and E. N. Fortson,
Search for a coupling of the Earth's gravitational field to nuclear spins in
atomic mercury, Phys. Rev. Lett. {\bf 68}, 135 (1992).

\bibitem{Gemmel}
C. Gemmel \emph{et al.},
Ultra-sensitive magnetometry based on free precession of nuclear spins,
Eur. Phys. J. D {\bf 57}, 303 (2010).

\bibitem{Gemmel2}
C. Gemmel \emph{et al.},
Limit on Lorentz and $CPT$ violation of the bound neutron
using a free precession ${}^3$He/${}^{129}$Xe comagnetometer,
Phys. Rev. D {\bf 82}, 111901(R) (2010).

\bibitem{Gemmel3}
M. Burghoff \emph{et al.}, Probing Lorentz invariance and other fundamental
symmetries in ${}^3$He/${}^{129}$Xe clock-comparison experiments,
J. Phys.: Conf. Ser. {\bf 295}, 012017 (2011).

\bibitem{Gemmel4}
W. Heil \emph{et al.}, Spin clocks: Probing fundamental symmetries in nature
Ann. Phys. (Berlin) {\bf 525}, 539 (2013).

\bibitem{Gemmel5}
F. Allmendinger, W. Heil, S. Karpuk, W. Kilian, A. Scharth, U. Schmidt,
A. Schnabel, Yu.~Sobolev, and K. Tullney,
New limit on Lorentz-invariance- and $CPT$-violating neutron spin
interactions using a free-spin-precession ${}^3$He-${}^{129}$Xe comagnetometer,
Phys. Rev. Lett. {\bf 112}, 110801 (2014).

\bibitem{HehlObukhovPuetzfeld}
F. W. Hehl, Y. N. Obukhov, and D. Puetzfeld, On Poincar\'{e} gauge theory
of gravity, its equations of motion, and Gravity Probe B,
Phys. Lett. A, \textbf{377}, 1775 (2013).

\bibitem{PRD}
A.~J. Silenko and O.~V. Teryaev, Semiclassical limit for Dirac
particles interacting with a gravitational field,
Phys. Rev. D {\bf 71}, 064016 (2005).

\bibitem{PRDfinal}
G. W. Bennett \emph{et al.} (Muon (\emph{g}-2) Collaboration),
Final report of the E821 muon anomalous magnetic moment measurement at BNL,
Phys. Rev. D \textbf{73}, 072003 (2006).

\bibitem{JPARC}
T. Mibe (for the J-PARC {\emph g}-2 collaboration),
{\emph New g-2 experiment at J-PARC}, Chinese Phys. C {\bf 34}, 
745-748 (2010).

\bibitem{MuEDM08}
G. W. Bennett {\emph et al. } (Muon ({\emph g}-2) Collaboration),
\emph{An improved limit on the muon electric dipole moment},
Phys. Rev. D {\bf 80}, 052008 (2009).

\bibitem{Melissinos}
A. C. Melissinos, {\it The Sagnac effect and the Tevatron},
arXiv:1101.4008 [physics.acc-ph].

\bibitem{Semertzidis}
Y. K. Semertzidis, {\it The Sagnac effect in the proton EDM experiment},
CAPP/IBS, Korea Advanced Institute of Science and Technology (KAIST), 
5 June 2014 (unpublished).

\bibitem{Dvor1}
M. Dvornikov, Neutrino interaction with matter in a noninertial frame,
JHEP {\bf 10}, 053 (2014).

\bibitem{Dvor2}
M. Dvornikov, Neutrino interaction with background matter in a non-inertial frame,
Mod. Phys. Lett. A {\bf 30}, 1530017 (2015).

\bibitem{KamTeryaev1}
A. Yu. Kamenshchik and O. V. Teryaev,
Chaotic spin precession in anisotropic universes and fermionic dark matter,
Phys. Part. Nucl. Lett. \textbf{13}, 298 (2016).

\bibitem{KamTeryaev2}
A. Yu. Kamenshchik and O. V. Teryaev,
Spin precession in anisotropic cosmologies,
Eur. Phys. J. C \textbf{76}, 293 (2016).

\end{thebibliography}
\end{document}